\documentclass[a4paper,10pt,twoside]{cpc-hepnp}
\usepackage{CJK,upgreek,fancyhdr}
\usepackage{multicol}
\usepackage{graphicx}
\usepackage{amssymb,bm,mathrsfs,bbm,amscd}
\usepackage[tbtags]{amsmath}
\usepackage{lastpage}
\usepackage{longtable}
\usepackage{changes}
\usepackage{etoolbox}
\usepackage{setspace}
\usepackage{tabularx}
\usepackage{lineno,hyperref}
\usepackage{supertabular}
\usepackage[normalem]{ulem}
\usepackage{eqnarray}
\usepackage{booktabs}  
\usepackage{subfigure}
\begin{document}
\begin{CJK*}{GB}{gbsn}

\fancyhead[c]{\small Chinese Physics C~~~Vol. xx, No. x (202x) xxxxxx}
\fancyfoot[C]{\small 010201-\thepage}

\title{New Geiger-Nuttall law for two-proton radioactivity
\thanks{We would like to thank X. -D. Sun, J. -G. Deng, J. -L. Chen and J. -H. Cheng for useful discussion. This work is supported in part by the National Natural Science Foundation of China (Grants No. 11205083, No.11505100 and No. 11705055), the Construct Program of the Key Discipline in Hunan Province, the Research Foundation of Education Bureau of Hunan Province, China (Grant No. 15A159 and No. 18A237), the Natural Science Foundation of Hunan Province, China (Grants No. 2015JJ3103, No. 2015JJ2121 and No. 2018JJ2321), the Innovation Group of Nuclear and Particle Physics in USC, the Shandong Province Natural Science Foundation, China (Grant No. ZR2015AQ007), the Hunan Provincial Innovation Foundation For Postgraduate (Grant No. CX20200909), the National Innovation Training Foundation of China (Grant No. 201910555161) and the Opening Project of Cooperative Innovation Center for Nuclear Fuel Cycle Technology and Equipment, University of South China (Grant No. 2019KFZ10).}}

\author{%
Hong-Ming Liu (ÁõºêÃú)$^{1}$
\quad You-Tian Zou (×ÞÓÐÌð)$^{1}$
\quad Xiao Pan (ÅËÏö) $^{1}$
\quad Jiu-Long Chen (³Â¾ÁÁú)$^{1}$
\quad Biao He (ºÎ±ë)$^{2}$\\
\quad Xiao-Hua Li (ÀîС»ª)$^{1,3,4;1)}$\email{lixiaohuaphysics@126.com}%
}
\maketitle
\address{%
$^1$ School of Nuclear Science and Technology, University of South China, Hengyang 421001, China\\
$^2$ College of Physics and Electronics, Central South University, 410083 Changsha, People¡¯s Republic of China\\
$^3$ Cooperative Innovation Center for Nuclear Fuel Cycle Technology \& Equipment, University of South China, Hengyang 421001, China\\
$^4$ Key Laboratory of Low Dimensional Quantum Structures and Quantum Control, Hunan Normal University, Changsha 410081, China\\
}
\begin{abstract}
In the present work, combining with the Geiger-Nuttall law, a two-parameter empirical formula is proposed to study the two-proton ($2p$) radioactivity. Using this formula, the calculated $2p$ radioactivity half-lives are in good agreement with the experimental data as well as the calculated ones obtained by Goncalves $et \ al.$ ([\href {10.1016/j.physletb.2017.09.032}{Phys. Lett. B \textbf{774}, 14 (2017)}]) using the effective liquid drop model (ELDM), Sreeja $et \ al.$ ([\href {10.1140/epja/i2019-12694-5}{Eur. Phys. J. A \textbf{55}, 33 (2019)}]) using a four-parameter empirical formula and Cui $et \ al.$ ([\href {https://doi.org/10.1103/PhysRevC.101.014301}{Phys. Rev. C  \textbf{101}: 014301 (2020)}]) using a generalized liquid drop model (GLDM). In addition, this two-parameter empirical formula is extended to predict the half-lives of 22 possible $2p$ radioactivity candidates, whose the $2p$ radioactivity released energy $Q_{2p} \textgreater 0$, obtained from the latest evaluated atomic mass table AME2016. The predicted results have good consistency with ones using other theoretical models such as the ELDM, GLDM and four-parameter empirical formula.
\end{abstract}


\begin{pacs}
{23.60.+e,} {21.10.Tg}
\end{pacs}

\footnotetext[0]{\hspace*{-3mm}\raisebox{0.3ex}{$\scriptstyle\copyright$}2020
Chinese Physical Society and the Institute of High Energy Physics
of the Chinese Academy of Sciences and the Institute
of Modern Physics of the Chinese Academy of Sciences and IOP Publishing Ltd}%
\begin{multicols}{2}
\section{Introduction}

In recent years, masses of works have been focused on the nuclei beyond the proton drip line due to the fact that they show new phenomena that can not  be found in the stable nuclei\cite{1,2,3}, e.g. the two-proton ($2p$) radioactivity phenomenon, which was predicted by Zel'dovich \cite{4} and Goldansky\cite{5,6} in the 1960s. From the perspective of pairing energy, the emitted two protons in $2p$ radioactivity process should be simultaneous emission from the ground state of a radioactive nucleus beyond the drip line. However, there is no agreement on the two protons are simultaneously emitted as two indepedent protons or ``diproton emission'' similar to the emission of a like-$^2$He cluster from mother nucleus. Obviously, for odd-proton-number (odd-$Z$) nuclei, proton radioactivity is the predominant decay mode, for even-proton-number (even-$Z$) nuclei lying near the proton drip line, may occur the $2p$ radioactivity phenomenon due to the effect of proton pairing \cite{7}. Experimentally, in 1978, the probability of $2p$ decay width of $^{16}$Ne and $^{12}$O was reported\cite{8}. Later, the ground-state true $2p$ radioactivity of $^{45}${Fe} was observed at Grand acc$\acute{\rm{e}}$$\rm{l}$$\acute{\rm{e}}$rateur national dions lourds (GANIL) \cite{9} and Gesellschaft f$\ddot{\rm{u}}$r Schwerionenforschung (GSI)\cite{10}, respectively. In 2005, the $2p$ radioactivity of $^{54}$Zn was detected at GANIL \cite{11}, followed by the $2p$ radioactivity of $^{48}$Ni was found \cite{12}. In 2007, the $2p$ radioactivity of $^{19}$Mg was revealed by tracking the decay products \cite{13}. Recently, the $2p$ emission of $^{67}$Kr was observed in an experiment with the BigRIPS separator\cite{14}.

Theoretically, up to now, there are various theoretical models, such as the direct decay model \cite{15,16,17,18,19,20,21}, the simultaneous versus sequential decay model \cite{22}, the diproton model \cite{23,24}, three-body model \cite{25,26,27,28},  have been proposed to investigate the $2p$ radioactivity. Using an $R$-matrix formula, B. A. Brown $et \ al.$ reproduced the $2p$ radioactivity half-lives of $^{45}$Fe\cite{29}, followed by using the continuum shell model, J. Rotureau $et \ al.$ microcosmically described the $2p$ radioactivity in $^{45}$Fe, $^{48}$Ni and $^{54}$Zn\cite{30}. In 2017,  M. Goncalves $et \ al.$ used the effective liquid drop model (ELDM) to calculate the half-lives of $2p$ radioactivity nuclei \cite{33}, their calculated results can reproduce the experimental data well\cite{31,32}. Meanwhile, based on the ELDM, they predicted the $2p$ radioactivity half-lives of 33 nuclei, whose the $2p$ radioactivity released energy $Q_{2p} > 0$, obtained from the latest evaluated atomic mass table AME2016 \cite{35,34}. In 2019, Sreeja $et \ al.$ proposed a four-parameter empirical formula to study the $2p$ radioactivity half-lives \cite{36}, these parameters are obtained by fitting the predicted results of Goncalves $et \ al.$ work \cite{33}. Their calculated results are in agree well with the known experimental data. Recently, Cui $et \ al.$ studied the $2p$ radioactivity of the ground state of nuclei based on a generalized liquid drop model (GLDM) \cite{37}, in which the $2p$ radioactivity process is described as a pair particle preformed near the surface of the parent nucleus penetrating the barrier between the cluster and daughter nucleus. In this view, $2p$ radioactivity shares the similar theory of barrier penetration with different kinds of charged particles' radioactivity, such as $\alpha$ decay, cluster radioactivity, proton radioactivity and so on\cite{38,39,41,42,43,40}. In our previous work\cite{44}, based on the Geiger-Nuttall (G-N) law\cite{45}, we proposed a two-parameter empirical formula of new G-N law for studying proton radioactivity, which can be treated as an effective tool to study proton radioactivity. Therefore, whether the G-N law can be extended to study the $2p$ radioactivity or not is an interesting topic. In this work, a two-parameter analytic formula, which is the relation of $2p$ radioactivity half-life $T_{1/2}$, $2p$ radioactivity released energy $Q_{2p}$, the charge of the daughter nucleus $Z_d$ and orbital angular momentum $l$ taken away by the two emitted protons, is proposed to study the $2p$ radioactivity.

This article is organized as follows. In next section, the theoretical framework for the new G-N law is described in detail. In Section \ref{section 3}, the detailed calculations, discussion and predictions are provided. In Section \ref{section 4}, a brief summary is given.

\section{Theoretical framework}
\label{section 2}
In 1911, Geiger and Nuttal found there is a phenomenological relationship between the $\alpha$ decay half-life $T_{1/2}$ and decay energy $Q_{\alpha}$. This relationship is so called as Geiger-Nuttal (G-N) law. It is expressed as
\begin{equation}
\rm{log_{10}}{\emph{T}}_{1/2} =  {\emph{a}}\,{Q_{\alpha}}^{-1/2} + {\emph{b}},
\end{equation} 
where $a$ and $b$ represent the two isotopic chain--dependent parameters of this formula. Later, the G-N law was widely applied to study the half-lives of $\alpha$ decay \cite{46,47,48,49}, cluster radioactivity \cite{50,51,52} and proton radioactivity \cite{53,54,55}. However, relative to $\alpha$ decay and cluster radioactivity, the proton radioactivity half-life is more sensitive to the centrifugal barrier, it leads to the linear relationship between the half-life of the proton radioactivity and the released energy $Q_p$ only exists for the proton radioactivity isotopes with the same orbital angular momentum $l$ taken away by the emitted proton\cite{53,55,44}. Similarly, the $2p$ radioactivity half-life maybe also strongly depends on the $2p$ radioactivity released energy $Q_{2p}$ and the orbital angular momentum $l$ taken away by the two emitted protons. Recently, considering the contributions of $Q_{2p}$ and the orbital angular momentum $l$ to the $2p$ radioactivity half-life, Sreeja $et \ al.$ put forward a four-parameter empirical formula to study the $2p$ radioactivity half-lives, which is expressed as \cite{36}
\begin{equation}
\rm{log_{10}}{\emph{T}}_{1/2} = (({\emph{a}}\times {\emph{l}}) + {\emph{b}})\,Z_{\emph{d}}^{\,0.8}\,{Q_{2\emph{p}}}^{-1/2} + (({\emph{c}}\times {\emph{l}}) + {\emph{d}}),
\label{subeq:2}
\end{equation}
where $a$ = 0.1578, $b$ = 1.9474, $c = -1.8795$ and $d =-24.847$ denote the adjustable parameters, which are obtained by fitting the calculated results of the ELDM \cite{33}. Their calculated results can reproduce the known experimental data well.

 In our previous work\cite{44}, considering the contributions of the daughter nuclear charge $Z_d$ and the orbital angular momentum $l$ taken away by the emitted proton to the proton radioactivity half-life, we proposed a two-parameter empirical formula of new G-N law for proton radioactivity. This formula is written as
\begin{equation}
\rm{log_{10}}{\emph{T}}_{1/2} = {\emph{a}_{\beta}}\,(Z_{\emph{d}}^{\,0.8}+{\emph{l}}^{\beta})\,{Q_{\emph{p}}}^{-1/2} + {\emph{b}_{\beta}},
\label{subeq:3}
\end{equation}
where $a_{\beta} = 0.843$ and $b_{\beta} = -27.194$ are the fitted parameters. The exponent on the orbital angular momentum $l$ taken away by the emitted proton, $\beta$ is 1, which is obtained by fitting 44 experimental data of the proton radioactivity in the ground state and isomeric state. Combined with Sreeja $et \ al.$ works \cite{36, 55} and our previous work \cite{44}, it is interesting to validate whether a two-parameter form of empirical formula is suitable to investigate the $2p$ radioactivity or not.  In this work, due to the experimental data of $2p$ radioactivity nuclei with orbital angular momentum $l \ne 0$ were not observed in experiments, we choose the experimental data of true $2p$ radioactivity nuclei ($^{19}$Mg, $^{45}$Fe, $^{48}$Ni, $^{54}$Zn and $^{67}$Kr) with $l$ = 0 and the predicted $2p$ radioactivity half-lives of 7 nuclei with $l \ne 0$ (1 case with $l$ = 1, 4 cases with $l$ = 2 and 2 cases with $l$ = 4) extracted from Goncalves $et \ al.$ work \cite{33} as the database.

At first, for the $\beta$ value describing the effect of $l$ on the $2p$ radioactivity half-life, we choose the $\beta$ value corresponding to the smallest standard deviation $\sigma$ between the database and the calculated $2p$ radioactivity half-lives as the optimal value, while the $\beta$ is varied from 0.1 to 0.5. The relationship between the $\sigma$ and $\beta$ value is shown in Fig. \ref {fig 1}. As we can see from this figure, it is clearly that the $\sigma$ is smallest when $\beta$ is equal to 0.25. Comparing with the $\beta$ value of Eq.\,(\ref{subeq:3}) reflecting the effect of $l$ on the proton radioactivity half-life, this $\beta$ value is smaller. The resaon may be the reduced mass $\mu$ of proton radioactivity nucleus is smaller than ones of $2p$ radioactivity nucleus, it leads to the contribution of centrifugal barrier on the half-life of $2p$ radioactivity nucleus is smaller. Correspondingly, the values of parameters $a$ and $b$ are given as
\begin{equation}
a = 2.032,\, b = 26.832,
\end{equation}
Then, we can obtain a final formula, which can be written as
\begin{equation}
\rm{log_{10}}{\emph{T}}_{1/2} = 2.032\,(Z_{\emph{d}}^{\,0.8}+{\emph{l}}^{\,0.25})\,{Q_{2\emph{p}}}^{-1/2} - 26.832.
\label{subeq:5}
\end{equation}
\begin{center}
\includegraphics[width=10.5cm]{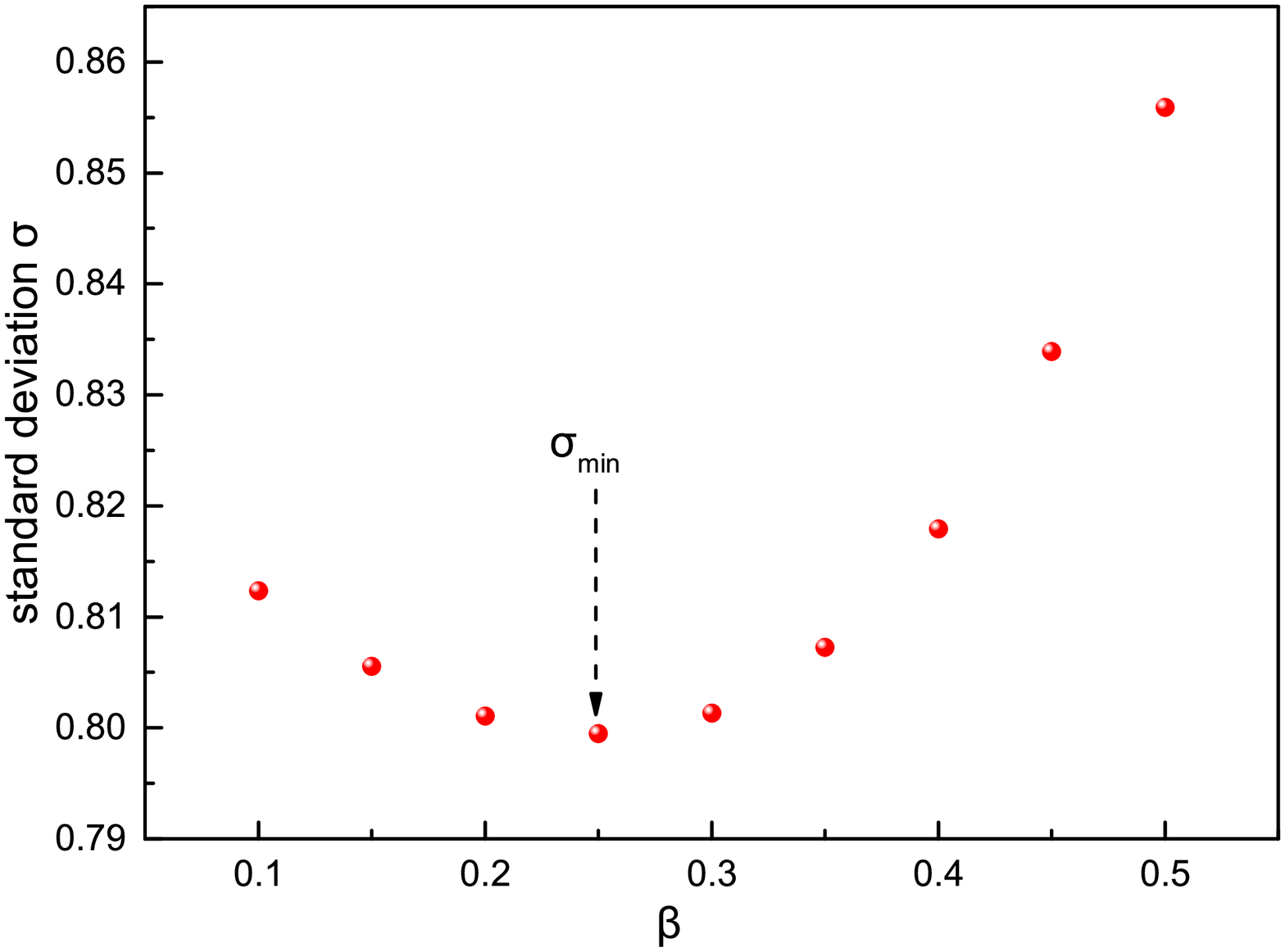}%Fig. 1
\figcaption{(color online) The releationship between the standard deviation $\sigma$ and the value of $\beta$.}
\label{fig 1}
\end{center}

\section{Results and discussion}
\label{section 3} 
The primary aim of this work is to verify the feasibility of using Eq.\,(\ref{subeq:5}) to investigate $2p$ radioactivity, while the calculated logarithmic half-lives of $2p$ radioactivity nuclei are listed in the seventh column of Table \ref{table 1}. Meanwhile, for comparison, the calculated results using GLDM, ELDM  and a four-parameter empirical formula are shown in the fourth to sixth column of this table, respectively. In Table \ref{table 1}, the first three columns denote the $2p$ radioactivity nucleus, the experimental $2p$ radioactivity released energy $Q_{2p}$ and the logarithmic experimental $2p$ radioactivity half-life ${\rm{log}}_{10}T_{1/2}^{\rm{exp}}$, respectively. For quantitative comparisons between the calculated $2p$ radioactivity half-lives using our empirical formula and experimental ones, the last column represents the logarithm of errors between the experimental $2p$ radioactivity half-lives and the calculated ones using our empirical formula $\rm{log_{10}}\emph{HF} = {\rm{log}}_{10}{{\emph{T}}_{1/2}^{\,\rm{exp}}} -{\rm{log}}_{10}{{\emph{T}}_{1/2}^{\,\rm{cal}}}$. From this table, it can be seen that for the true $2p$ radioactivity nuclei $^{19}$Mg, $^{45}$Fe, $^{48}$Ni, $^{54}$Zn and $^{67}$Kr ($Q_p < 0$, $Q_{2p} > 0$ ), most values of $\rm{log_{10}}\emph{HF}$ are between -1 and 1. Particularly, for the cases of $^{48}$Ni of $Q_{2p}$ =  1.290 and $^{45}$Fe of $Q_{2p}$ = 1.154, the values of $\rm{log_{10}}\emph{HF}$ are 0.07 and 0.24, indicating our calculated results can reproduce the experimental data well. As for the not true $2p$ radioactivity nuclei $^{6}$Be, $^{12}$O and $^{16}$Ne ($Q_p > 0$, $Q_{2p} > 0$), the values of $\rm{log_{10}}\emph{HF}$ for $^{6}$Be and $^{16}$Ne are relatively large. Likewise, the differences between the calculated $2p$ radioactivity half-lives using GLDM, ELDM and a four-parameter empirical formula and the experimental data are more than three orders of magnitude. The reason may be limited by the early experimental equipments, the measured decay widths of these $2p$ radioactivity nuclei were not accurate enough. To measure the experimental $2p$ half-lives of these nuclei again is meaningful in the future. In the case of $^{12}$O, the values of $\rm{log_{10}}\emph{HF}$ are small, impling our formula maybe also suitable for studing the not true $2p$ radioactivity nuclei having relatively accurate experimental data.

\begin{center}
\includegraphics[width=10.5cm]{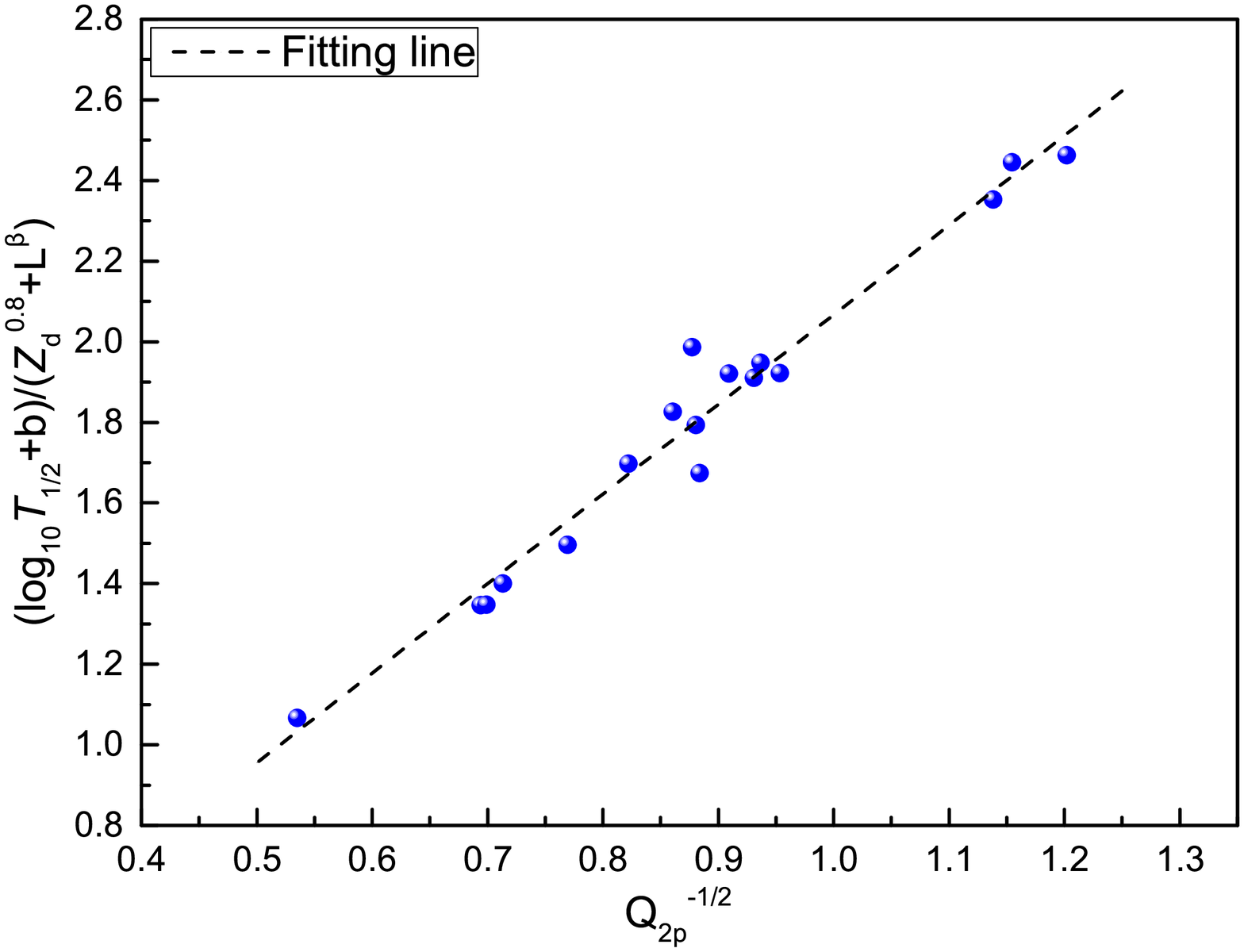}%Fig. 2
\figcaption{(color online) The linear relationship between the quantity $[\rm{log_{10}}{\emph{T}}_{1/2} + 26.832]/(Z_{d}^{0.8}+\emph{l}^{\,0.25})$ and $Q_{2p}$ for the database used to fit the parameters of Eq.\,(\ref{subeq:5}).}
\label{fig 3}
\end{center}
\end{multicols}

\begin{table*}[!hbt]%Table 1
\centering
\caption{The comparison of the experimental data of the $2p$ radioactivity nuclei with different theoretical models (GLDM, ELDM, the four-parameter empirical formula of Ref. \cite{36} and our empirical formula). The experimental data are taken from the corresponding references.}
\label{table 1}
\renewcommand\arraystretch{1.5}
\setlength{\tabcolsep}{2.5pt}
\setlength\LTleft{-13in}
\setlength\LTright{-13in plus 1 fill}
\begin{tabular}{ccccccccc}
\hline\noalign{\smallskip}
\hline\noalign{\smallskip}
 Nucleus & $Q_{2p}^{\rm{\,exp}}$ (MeV) &$\rm{log_{10}}{\emph{T}}_{1/2}^{\,exp}$ (s) &$\rm{log_{10}}{\emph{T}}_{1/2}^{\,GLDM}$(s)\cite{37}&  $\rm{log_{10}}{\emph{T}}_{1/2}^{\,ELDM}$(s)\cite{33}& $\rm{log_{10}}{\emph{T}}_{1/2}^{}$ (s) \cite{36}& $\rm{log_{10}}{\emph{T}}_{1/2}^{\,This\;work}$ (s)&$\rm{log_{10}}$\emph{HF}\\
\noalign{\smallskip}\hline\noalign{\smallskip}									
$^{6}$Be	&	1.371\cite{56}	&	$-20.30$\cite{56} 	&	$-19.37$ 	&	$-19.97$ 	&	$-21.95$ 	&	$-23.81$	&	3.51\\
	\hline
$^{12}$O	&	1.638\cite{57}	&	$\textgreater-20.20$\cite{57} 	&	$-19.17$ 	&	$-18.27$ 	&	$-18.47$ 	&	$-20.17$&	$\textgreater-0.03$ 	\\
&	1.820\cite{8}	&	$-20.94$\cite{8} 	&	$-20.94$ 	&--		&$-18.79$		&	$-20.52$&	$-0.42$\\
	&	1.790\cite{58}	&	$-20.10$\cite{58} 	&	$-20.10$ 	&	--	&	$-18.74$	&$-20.46$&$0.36$\\
	&	1.800\cite{59}	&	$-20.12$\cite{59} 	&	$-20.12$ 	&	--	&$-18.76$		&$-20.48$&$0.36$\\
	\hline
$^{16}$Ne	&	1.33\cite{8}	&	$-20.64$\cite{8} 	&	$-16.45$ 	&	--	&	$-15.94$ 	&	$-17.53$&	$-3.11$\\
	&	1.400\cite{60}	&	$-20.38$\cite{60} 	&	$-16.63$ 	&	$-16.60$ 	&	$-16.16$ 	&	$-17.77$&$-2.61$\\
		\hline
$^{19}$Mg	&	0.750\cite{13}	&	$-11.40$\cite{13} &$-11.79$ 	&	$-11.72$ 	&	$-10.66$ 	&	$-12.03$&	$0.63$\\
	\hline
$^{45}$Fe	&	1.100\cite{10}	&	$-2.40$\cite{10} 	&	$-2.23$ 	&	--	&	$-1.25$	&	$-2.21$&$-0.19$\\
	&	1.140\cite{9}	&	$-2.07$\cite{9} 	&	$-2.71$ 	&	--	&	$-1.66$ 	&	$-2.64$&$0.57$\\
	&	1.210\cite{61}	&	$-2.42$\cite{61} 	&	$-3.50 $	&	--	&	$-2.34$ 	&$-3.35$&$0.93$\\
	&	1.154\cite{12}	&	$-2.55$\cite{12} 	&	$-2.87$ 	&	$-2.43$ 	&	$-1.81$ 	&	$-2.79$&$0.24$\\
		\hline
$^{48}$Ni	&	1.350\cite{12}	&	$-2.08$\cite{12} 	&	$-3.24$ 	&	--	&	$-2.13$ 	&$-3.13$&$1.05$\\
	&	1.290\cite{62}	&	$-2.52$\cite{62} 	&	$-2.62 $	&	--	&	$-1.61$ 	&	$-2.59$	&	$0.07$\\
	\hline
$^{54}$Zn	&	1.480\cite{63}	&	$-2.43$ \cite{63}	&$-2.95$ 	&$-2.52$ 	&	$-1.83$ 	&$-2.81$&$0.38$	\\
	&	1.280\cite{64}	&	$-2.76$\cite{64} 	&	$-0.87$ 	&	--	&	$-0.10$ 	&	$-1.01$&$-1.75$	\\
		\hline
$^{67}$Kr	&	1.690\cite{14}	&	$-1.70$\cite{14}	&	$-1.25$ 	&	$-0.06$ 	&	0.31 	&	$-0.58$&$-1.12$\\
\noalign{\smallskip}\hline
\noalign{\smallskip}\hline
\end{tabular}
\end{table*}

\begin{table*}[!hbt]%Table 2
\centering
\caption{The comparison of calculated $2p$ radioactivity half-lives using GLDM, ELDM, the four-parameter empirical formula from Ref. \cite{36} and our empirical formula. The $2p$ radioactivity released energy $Q_{2p}$ and orbital angular momentum $l$ taken away by the two emitted protons are taken from Ref. \cite{33}.}
\label{table 2}
\renewcommand\arraystretch{1.5}
\setlength{\tabcolsep}{3pt}
\setlength\LTleft{-10in}
\setlength\LTright{-10in plus 1 fill}
\begin{tabular}{ccccccc}
\hline\noalign{\smallskip}
\hline\noalign{\smallskip}
Nucleus &$Q_{2p}$ (MeV)&$l$  &$\rm{log_{10}}{\emph{T}}_{1/2}^{\,GLDM}$ (s)\cite{37}& $\rm{log_{10}}{\emph{T}}_{1/2}^{\,ELDM}$(s)\cite{33} &  $\rm{log_{10}}{\emph{T}}_{1/2}$(s)\cite{36} & $\rm{log_{10}}{\emph{T}}_{1/2}^{\,This\;work}$ (s)\\
\noalign{\smallskip}\hline\noalign{\smallskip}
$^{22}$Si	&	1.283	&	0	&$-13.30$&$-13.32$&$-12.30$&$-13.74$\\
$^{26}$S	&	1.755	&	0	&$-14.59$&$-13.86$&$-12.71$&$-14.16 $\\
$^{34}$Ca	&	1.474	&	0	&$-10.71$&$-9.91$&$-8.65$&$-9.93$\\
$^{36}$Sc	&	1.993	&	0	&$		$&$-11.74$&$-10.30$&$-11.66$\\
$^{38}$Ti	&	2.743	&	0	&$-14.27$&$-13.56$&$-11.93$&$-13.35$\\
$^{39}$Ti	&	0.758	&	0	&$-1.34$&$-0.81$&$-0.28$&$-1.19$\\
$^{40}$V	&	1.842	&	0	&$	$&$-9.85$&$-8.46$&$-9.73$\\
$^{42}$Cr	&	1.002	&	0	&$-2.88$&$-2.43$&$-1.78$&$-2.76$\\
$^{47}$Co	&	1.042	&	0	&$		$&$-0.11$&$0.21$&$-0.69$\\
$^{49}$Ni	&	0.492	&	0	&$	14.46 	$&$14.64$&$12.78$&$12.43$\\
$^{56}$Ga	&	2.443	&	0	&$		$&$-8.00$&$-6.42$&$-7.61$\\
$^{58}$Ge	&	3.732	&	0	&$-13.10$&$-11.74$&$-9.53$&$-10.85$\\
$^{59}$Ge	&	2.102	&	0	&$-6.97$&$-5.71$&$-4.44$&$-5.54$\\
$^{60}$Ge	&	0.631	&	0	&$	13.55 	$&$	14.62 	$&$	12.40 	$&$12.04$\\
$^{61}$As	&	2.282	&	0	&$		$&$-6.12$&$-4.74$&$-5.85$\\
\hline
$^{10}$N	&	1.3	&	1	&$		$&$-17.64$&$-20.04$&$-18.59$\\
$^{28}$Cl	&	1.965	&	2	&$		$&$-12.95$&$-14.52$&$-12.46$\\
$^{32}$K	&	2.077	&	2	&$		$&$-12.25$&$-13.46$&$-11.55$\\
$^{57}$Ga	&	2.047	&	2	&$		$&$-5.30$&$-5.22$&$-4.14$\\
$^{62}$As	&	0.692	&	2	&$		$&$	14.52 	$&$13.83$&14.18\\
$^{52}$Cu	&	0.772	&	4	&$		$&$	9.36 	$ &$8.62$&8.74\\
$^{60}$As	&	3.492	&	4	&$		$ & $-8.68$ & $-10.84$ & $-8.33$\\
\noalign{\smallskip}\hline
\noalign{\smallskip}\hline
\end{tabular}
\end{table*}
%%%
\begin{multicols}{2}
To further test the feasibility of our empirical formula, using Eq.\,(\ref{subeq:5}), we also predict the $2p$ radioactivity half-lives of 22 nuclei, whose the $2p$ radioactivity released energy $Q_{2p} \textgreater 0$. The $Q_{2p}$ values are taken from the latest evaluated atomic mass table AME2016 and shown in the second column of Table \ref{table 2}. In this table, the first and third columns represent the $2p$ radioactivity candidates, the angular momentum $l$ taken away by the two emitted protons, respectively. For benchmark, the predicted results using GLDM, ELDM and the four-parameter empirical formula, which are extracted from Ref.\;\cite{37}, \cite{33} and \cite{36}, are also listed in this table. From this table, we can clearly see that for the cases of $l\ne 0$, the predicted results using our empirical formula are more close to the predicted ones using ELDM comparing with the predicted  results using the four-parameter empirical formula, most of the predicted results are of the same order of magnitude. As an example, in the cases of $^{28}$Cl ($^{60}$As), the predicted $2p$ radioactivity half-lives using ELDM, the four-parameter empirical formula and our empirical formula are $-12.95$ ($-8.68$), $-14.52$ ($-10.84$) and $-12.46$ ($-8.33$), respectively. It implies our empirical formla is also suitable for study the nuclei with orbital angular momentum $l\neq0$. In the cases of $l = 0$, the predicted $2p$ radioactivity half-lives using our empirical formula are in good agreement with the ones of GLDM and ELDM. For further demonstrating the significant correlation between the $2p$ radioactivity half-lives $T_{1/2}$ and the $2p$ radioactivity released energies $Q_{2p}$, based on Eq.\,(\ref{subeq:5}), we plot the quantity $[\rm{log_{10}}{\emph{T}}_{1/2} + 26.832]/(Z_{d}^{0.8}+\emph{l}^{\,0.25})$ as a function of $Q_{2p}^{-1/2}$ in Fig. \ref{fig 3}. In this figure, there is an obvious linear dependence of $\rm{log}_{10}\emph{T}_{1/2}$ on $Q_{2p}$$^{-1/2}$ while removing the contributions of charge number $Z_{d}$ and orbital angular momentum $l$ on the $2p$ radioactivity half-lives. 
\end{multicols}
 
\begin{multicols}{2}
\section{Summary}
\label{section 4}
In this work, considering the contributions of the charge of daugher nucleus $Z_{d}$ and the orbital angular momentum $l$ taken away by the two emitted protons, a two-parameter empirical formula of new Geiger-Nuttall law is proposed for studying the $2p$ radioactivity. Using this formula, the experimental data of the true $2p$ radioactivity nuclei can be reproduced well. Meanwhile, it is found that the calculated results using our empirical formula are agreement with ones of GLDM, ELDM and the four-parameter empirical formula. Moreover, using our formula, the half-lives of possible $2p$ radioactivity candidates are predicted. The predicted results may be provide theoretical help for the future experiments.
\end{multicols}
\vspace{-1mm}
\centerline{\rule{80mm}{0.1pt}}
\vspace{2mm}

\begin{multicols}{2}

\end{multicols}

\clearpage
\end{CJK*}
\end{document}